\documentclass[12pt,A4]{article}
\usepackage{epsfig}
\usepackage{array}
\usepackage{amsmath}
\usepackage{amssymb}

\newcommand{\be}{\begin{equation}}
\newcommand{\ee}{\end{equation}}
\newcommand{\bea}{\begin{eqnarray}}
\newcommand{\eea}{\end{eqnarray}}

\begin{document}

\begin{center}
{\large\bf THE CRITICAL NOTES TO THE SOLUTIONS OF THE ``PUZZLE'' OF
HYPERFINE STRUCTURE  IN 82+ AND 80+ $^{209}$BI IONS}

\vspace*{2cm}
{ F. F. Karpeshin \\ Mendeleev All-Russian Research
Institute of Metrology \\ 190005 Saint-Petersburg, Russia \\
\bigskip
and \\ M. B. Trzhaskovskaya \\
Petersburg Nuclear Physics Institute
of the National Research Center ``Kurchatov Institute'' \\
198300 Gatchina, Russia }
\end{center}

\vspace*{3cm}
\abstract{ Some aspects of description of the Bohr---Weisskopf effect in hyperfine splitting of the H-- and Li--like ions of $^{209}$Bi are considered by application of the surface and volume models of the nuclear currents. Extension of these models, used in internal conversion theory, to description of the HFS allows one to successfully describe the effect, without resorting to the specific differences. The latters  are shown not to be needed at all. Moreover, they turn out to depend on the nuclear model even stronger than the HFS values themselves. Comparison of the calculated HFS values to  experiment shows a satisfactory agreement. Both models provide equally good description of the effect. However, they result in different values of the retrieved rms radius of the nuclear magnetization. In this respect, situation resembles the proton radius puzzle. Prospects of future research are discussed.
}

\vspace*{1cm}

\newpage

\large

\section{Introduction}

       A considerable progress during the past decades was archived in investigation of few-electron heavy ions. Specifically, this concerns study of their electronic structure and its influence on the nuclear processes. Wonderful experiments were performed studying the shell effects on the beta decay \cite{beta}. There is the comparative study of $\alpha$ decay in H-like and He-like ions on the urgent agenda, with respect to that in neutral atoms \cite{alf1,alf2,alf3}. In spite of  that the influence of the electron screening on the $\alpha$ decay is a very important question, in view of many applications in astrophysics and experiments with laser-produced plasma, it is only recently that the adequate approach has been found \cite{alf1}. It was clearly shown that the frozen-shell approximation, which was used during half century, exaggerates the effect of the shell at least  by an order of magnitude \cite{alf2,alf3}. Moreover, it gives the wrong sign of the effect \cite{alf3}. Furthermore, attractive ideas concern possibilities of manipulations by the electromagnetic decay of the nuclei. A way of drastic acceleration of nuclear decay rate by means of resonance conversion was proposed by \cite{kabzon,chin1,book}.

    A considerable attention was paid to study of the hyperfine structure of heavy ions.   As compared to neutral atoms, a few-electron wavefunction can be calculated with high accuracy in heavy ions.  On the other hand, QED effects give a significant contribution. This gave basis to suggest that the hyperfine splitting (HFS) can be used  to test QED (e.g. \cite{shab1} and Refs. cited therein). However, it was noted \cite{shab1} that
    there is a stumbling stone on the way  represented by
the Bohr---Weisskopf effect \cite{BW}. This effect generates known for decades hyperfine magnetic anomalies in  optical spectra of atoms. The effect is caused by the interaction of atomic electrons with the spatially-distributed magnetic moment of the nucleus. Although its contributions to the HFS of the $1s$ and $2s$ levels in the H-like and Li-like $^{209}$Bi ions comprise approximately 2\% and 2.2\%, respectively,  its actual contribution   depends on the nuclear model.
Some attempts were undertaken, aimed at calculation of this effect (\cite{shab1} and Refs. cited therein). They showed, however, that there remains a contradiction with theory at the level of 20 -- 30 percent. Such a result should be expected, because nuclear calculations still cannot be performed {\it ab initio} in principle, in view of absence of an appropriate parameter \cite{mig}. At the same time, the
Bohr---Weisskopf
effect becomes essential for description  of experimental data \cite{Lochp}.  In view of this problem, another, roundabout way was proposed
in Ref. \cite{shab1}. It runs that, instead of calculation of the
Bohr---Weisskopf effect, one can cancel its contribution in the specially constructed linear combination (difference) $\Delta'E$  (see Eq. (\ref{sde}) in section \ref{formul}) of the HFS values of the H- and Li-like ions, being in the $1s$- and $2s$ states, respectively. The cancellation supposedly takes place if the parameter $\zeta$ in the combination (see Eq. (\ref{zet}) in section \ref{formul}) is calculated in such a way that the BOHR---WEISSKOPF contribution is mutually subtracted in the difference. This combination was called specific difference (SD). In the case of $^{209}$Bi ions, the calculated value of $\zeta$ = 0.16886 was then listed to fifth decimal \cite{Lochp,shab2,Loch}.

       The first thing that catches your eye   is that consideration of such defined SD is in contradiction with general methods of epistemology, which is founded on comparison of experimental data with  theoretical values of the observables. But SD, thus defined, cannot be observed experimentally.  Moreover, the operation of subtraction, aimed at mutual cancellation of small terms, is incorrect from the viewpoint of mathematics as it leads also to a considerable reduction of the main parts. We note in this relation that uncertainties do not cancel one another in subtraction. Oppositely, they are summarized in the general case. This makes the result of subtraction, that is SD in our case, less accurate than  the $1s$- and $2s$ HFS values themselves. We will see this in Section \ref{res} and Table \ref{tfin}.

    And even more: for the method to work, it is necessary that parameter $\zeta$ be model independent. This idea seems to be absolutely incorrect, judging by experience of application of the  theory of internal conversion (IC). It
was subject to critical check in Refs. \cite{NP,yaf}. At first sight, application of IC theory may seem unusual. However, this idea is not new. For the first time, IC theory was applied in Ref. \cite{reiner}. Then estimations of the HFS values and the dynamic effect were performed in Refs. \cite{book,HF,echa}.
In the next section, we will show the relatedness of these two phenomena in more detail. Meanwhile, based on the methods of IC theory, the method of the magnetic moments was developed in Ref. \cite{NP} for interpretation of the penetration effects of the nuclear structure.
The method was then applied in Ref. \cite{yaf}, where it was shown that the conclusion of the ``hyperfine puzzle'' \cite{new} was mostly due to underestimation of the model dependence of the SD. Actually, there was no puzzle at that moment,  the authors should merely cite the result of paper \cite{NP}, where encounter with such a ``puzzle'' was literally predicted. Herein, we attack the problem in a different way as compared to \cite{yaf}. We study the penetration effect on the HFS and SD values by means of comparison of the two conventional models, which are known to work well in the IC theory: surface (SC) \cite{sliv} and volume (VC) \cite{book,muon} nuclear currents. These models  are opposite to one another in their physical sense. Due to the latter circumstance, the results obtained  can be considered as quite general.

       In section \ref{formul}, we derive shortly the formulas. The results of the calculations are reported in Section \ref{res}.
Unexpectedly, we arrived at the conclusion that both of the models work equally well in description of the data at the present level of precision. With different parameters, the models give the same values of the HFS for the $1s$- and $2s$ levels up to six decimals.
However, the HFS values are, naturally,  very sensitive to the only parameters of the models --- their radii. The $\zeta$ and $\Delta'E$ values turn out to be  more sensitive, than the HFS values themselves, as expected.
The results are discussed in more detail in the conclusive section. In the same section, we consider analogy and possible applications of the results for better understanding of  the proton radius puzzle.

    When the manuscript was ready, paper \cite{new} was issued. The authors show that the value of the magnetic moment of the nucleus $\mu$ =4.092 nuclear magnetons may be more correct than 4.1106 \cite{lederer} used previously. Such a new value would essentially keep the present results, merely rescaling them by
$\sim 0.5$\%. For the sake of completeness, the rescaled values are also presented. They demonstrate absence of any ``hyperfine puzzle'' within the present scope of the
$^{209}$Bi issue: the data can be fairly explained with any of the magnetic moment values.

\section{Remind of the model}
\label{formul}

\begin{figure}[!tb]
\centerline{ \epsfxsize=10cm\epsfbox{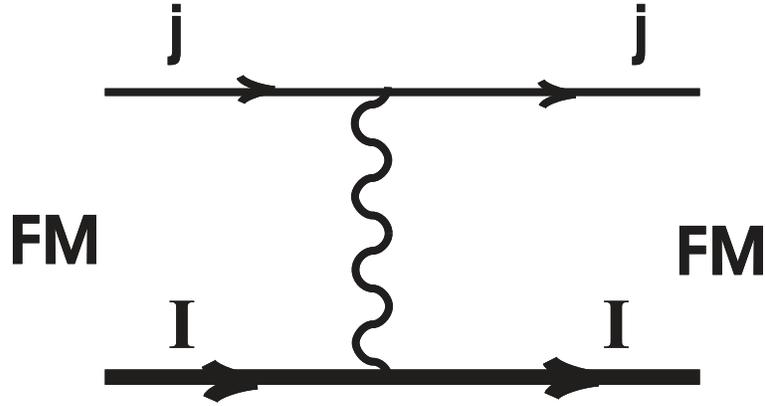}}
\caption{\footnotesize Feynman graph of hyperfine shift. Nuclear propagator is shown by bold line. Atomic state is defined by the total momentum $F$ and its projection $M$, together with $I$ and $j$ --- nuclear and electronic spins, respectively.}
\label{figHFS}
\end{figure}

\begin{figure}[!bt]
\centerline{ \epsfxsize=10cm\epsfbox{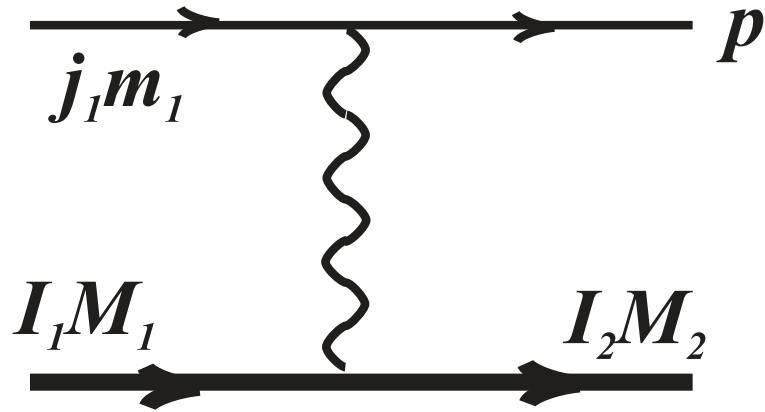}}
\caption{\footnotesize Feynman graph of internal conversion. $I_1$, $M_1$, $I_2$, $M_2$ --- nuclear quantum numbers (spins and their projections on the quantization axes) in the initial and final states, respectively. $j_1$, $m_1$ --- electronic quantum numbers in the initial state. Conversion electron is characterized with the four-vector of its momentum $p$.}
\label{figIC}
\end{figure}

       Feynman graphs of the hyperfine splitting and internal conversion are presented in Figures \ref{figHFS} and \ref{figIC}, respectively. Actually, the both graphs describe the same amplitude, though defined on different areas of the external kinematical variables of the transition energy and angular momenta. In quantum mechanics and theory of field, such  values can be related to each other by making use of the analytical properties of the amplitudes, and the processes themselves are spoken about as crossing channels.  The method of complex transition trajectories  by Landau (e. g. \cite{lan3,cmplx}), or complex angular momenta by Regge \cite{regge} may set examples. In the case of HFS, the analyticity of the amplitudes in Figs. \ref{figHFS} and \ref{figIC} means that all the methods, developed in the IC theory, can be directly applied to description of HFS values,  considered as the diagonal IC matrix elements in the limit of the transition energy $\omega \to 0$.

      The diagram in Fig. \ref{figIC} hints (correctly) that the amplitude of conversion transition factorizes into the amplitudes of virtual photon emission and its subsequent absorption by an atomic electron. As a consequence, internal conversion coefficient (ICC) is defined as the ratio of the probabilities
of the conversion $\Gamma_c(\tau, L)$ and radiative $\Gamma_\gamma (\tau, L)$ transitions:
\be
\alpha(\tau, L)  = \Gamma_c(\tau, L) / \Gamma_\gamma (\tau, L)\,,
\ee
where $\tau,  L$ stand for the type and multipole order of the transition.
We remind that ICC, as well as  HFS, would be independent of the nuclear model in the limit of the point-like nuclei,  where the electron penetration effects into the nuclear volume are absent. This approximation comprises the  ``no-penetration'' (NP) nuclear model. Calculations within the framework  of realistic nuclear models generally needs a two-dimensional  integration     over the electronic and nuclear variables  within the nuclear volume (e. g. \cite{NP}). This  results in loss of the factorization, which makes ICC  $\alpha(\tau, L) $ model-dependant, as well as HFS. Some reasonable models, which take into account the penetration effects, with also keeping the amplitudes  factorized,  were  introduced and approved by comparison to experiment. The SC nuclear model served as a basis for a number of tables of ICC (e.g. \cite{RAINE,RAI3}), highly demanded for the research and application purposes. The VC model is expected to work even better in the case of the valence $h_{9/2}$ proton orbital in $^{209}$Bi. Furthermore, the VC model was applied for description of muonic conversion \cite{book, muon,star}.

   Generally, in the IC theory, manifestation of the nuclear structure is classified into two kinds: static and dynamical effects (\cite{sliv, RAINE,RAI3,church,anom,hardy,raman} and Refs. cited therein). To the first kind belong the effects that arise because of change in the   electronic wave functions as compared to the Dirac wavefunctions  for the point-like nucleus. Coulomb wavefunctions are singular at the origin. Accounting for  the finite charge distribution over the nuclear  volume makes the functions regular, and brings about a correction to the internal conversion coefficients (ICC) up to 30 percent in the case of the $M1$ transitions in heavy nuclei \cite{RAINE,church}.
Turning now to paper \cite{shab1}, we note that it is directly pointed out therein that holdness of the $\zeta$ and $\Delta'E$ [��1]values was checked against    variation of the parameters of Fermi charge distribution over the nuclear volume. This just comprises the static effect. Therefore, the conclusion of the model independence of SD \cite{shab1} is only drawn from investigation of the static effect, which is known not to be essential after the main shortcoming --- the divergence of the Coulomb wavefunction --- is resolved, indeed.

       Influence of a model, used for the transition nuclear currents on the ICC values, is called the dynamical effect of the nuclear structure. It is just the dynamical effect which is responsible for differences between the experimental ICC and their table values, which are observed in some cases of forbidden nuclear transitions. These differences are also called anomalies in IC, similar to magnetic anomalies in the hyperfine spectra (e.g., \cite{kara,barza,pirs}).
    The dynamical effect constitutes up to  $\sim$10 percent in  heavy nuclei in the case of the $M1$ transitions. Turning now to HFS, we note that there is no transitions here. It is the distribution of the magnetism over the nuclear volume instead, which brings about the dynamical  effect.

    To a certain extent, the dynamical effect on the HFS values in the Li-like ions was tested in Refs. \cite{soff,chen}, using two very close to one another nuclear models (see below). But not on the SD's, which concept was proposed later. Actually, it was only in \cite{NP} that the SD values, calculated in  different nuclear models, were compared to one another for the first time. The difference discovered comprised three percent for the SD value, which is quite a crucial value.

    Within the framework of the IC theory, a general expression for HFS, allowing for the Bohr---Weisskopf effect, was obtained in Ref. \cite{NP}. Resulting expression for HFS of an electronic level reads as follows:
\bea
W=Nw\,, \nonumber \\
 w=\int_0^\infty g(r)f(r)dr +t^\nu \equiv w_0+t^{\nu} \,, \label{BW}\\
N=-\frac{2(2I+1)}{I(j+1)}e\kappa\mu \frac{e\hbar}{2M_pc} \,. \nonumber
\eea
Here $g(r)$, $f(r)$ are the large and small components of the radial Dirac electronic wavefunction of the $i$-th level.   $\kappa$ is the relativistic quantum number;
$j$, $I$ --- the electronic and nuclear spins, respectively,
$e$ --- the elementary charge, $\mu$ --- the magnetic moment of the nucleus, and $\frac{e\hbar}{2M_pc}$ --- the nuclear magneton.
$w_0$ gives the NP value, $t^\nu$, which we shall call the penetration matrix element,  bears information on the nuclear structure. Therefore, the $t^\nu$ value depends on the nuclear model. In the SC and VC nuclear models,
\be
t^{\nu}= \int_0^{R_c} g(r)f(r) Y^\nu(r)\ r^2 dr \,, \label{BWt}
\ee
with
\bea
\raisebox{-1em}{$Y^\nu(r)$\ = \ } &  \frac r{R_c^3}-\frac1{r^2} &
                                              \text{for } \nu=\text{SC}  \label{sc}\\
     &  \frac1{R_c^3}\left( 4r-3\frac{r^2}{R_c}\right)-\frac1{r^2}
     & \text{for } \nu=\text{VC}  \label{vc}
\eea
In the NP model $Y^{\nu}(r)\equiv 0$. $R_c$ is the model radius of the transition currents. We refer upper index $\nu$ to the model, and lower index $i$ --- to the electronic level.

       It is thus $t^{\nu}$ which only bears information about the
Bohr---Weisskopf effect. It was proposed to get rid off  it in the linear combination, called specific difference:
\be
\Delta' E= W_{2s}-\zeta W_{1s}\,.  \label{sde}
\ee
In terms of the penetration matrix elements (\ref{BW}), Eq. (\ref{sde}) has an evident solution
\be
\zeta=t_{2s}^{\nu}/t_{1s}^{\nu} \,. \label{zet}
\ee
By making use of the last equation, SD (\ref{sde}) can be also expressed in equivalent form as follows:
\bea
\Delta' E= Nw_0^{2s}-\zeta w_0^{1s} = \label{sd0}  \\
= Nw^{2s}\frac{p_{1s}-p_{2s}}{p_{1s}} \,, \label{sda}
\eea
 where
$p^\nu_i = {t^\nu_i}/{w^\nu_i}$ --- the relative contribution of the
Bohr---Weisskopf effect to the HFS. In view of that $p_{2s}>p_{1s}$, it follows from Eq. (\ref{sda}) that $\Delta'E <0$.

       Expression (\ref{BW}) is equivalent to those, used in Refs. \cite{soff,chen}, and which are based on the classical analogies. Moreover,
(\ref{sc}) for the SC model was used in papers \cite{soff,chen} under the name of ``homogenius distribution''.
In terms of Eq. (\ref{sc}), model \cite{soff,chen} is
reproduced by replacement of $Y(r)$ with
\be \tilde Y^n(r) =
\frac{r^{n+1}}{{R_c}^{n+3}} - \frac1{r^2}\,.  \label{Cheneq}
\ee
Model (\ref{Cheneq}) with parameter $n$ = 0, 2 was used in Refs. \cite{soff,chen}. In a particular case of $n$ = 0, Eq.
(\ref{Cheneq}) coincides with (\ref{sc}).  This seeming paradox
with the name has a simple explanation on the physical ground. In
fact, it is not the transition density, but rather the transition
current which determines the HFS ({\it e. g.} \cite{NP}). On the other hand,  such a classical picture of ``homogenius'' density distribution of elementary  point-like magnetic
dipoles leads to appearance of the effective current along the circle surrounding the locus, as
inside the locus all the elementary currents mutually cancel one another.
This directly generates the $\delta$-form nuclear current, as
\begin{figure} \centerline{ \epsfxsize=10cm\epsfbox{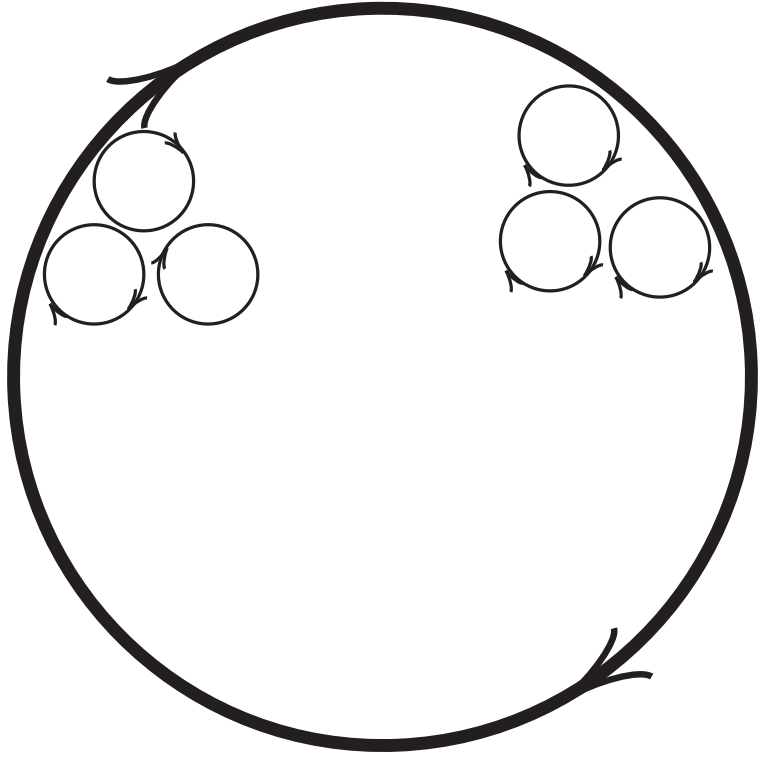}}
\caption{\footnotesize Schematic picture explaining seeming
contradiction of the terms of ``homogeneous'' and
``surface-current'' for the same nuclear model. Homogeneously
distributed inside the nucleus magnetic dipoles are shown by
elementary circular currents. In the bulk, the adjacent currents
mutually cancel one another. As a result, only the encircling
effective current survives, resulting in the surface-current
model. It is this surface current which brings about the hyperfine
splitting. } \label{scf}
\end{figure}
illustrated in Fig. \ref{scf}. It is worthy of noting that another such paradox was shown in Ref. \cite{star}. It was found that muonic conversion from the states of giant dipole resonance is better described by the VC model, in spite of that the transition nuclear density has a sharp maximum on the nuclear surface.

    It is worthy of noting some relations concerning the physical sense of the model parameters. In the VC model, $R_c$, like the equivalent electromagnetic radius, equals the radius of the sphere with the homogeneos sharp edge distribution of the magnetization currents. It is related with   the rms radius
$R_2$ by means of the following expression:
\be
R_2^{VC}=\sqrt{\frac23} R_c^{VC}\,.
\ee
In the SC model, all the multipole moment radii equal $R_i^{SC}\equiv R_c^{SC}$.

\section{Results of calculations}
\label{res}

As it was pointed out in the original paper by Bohr and Weisskopf,  the effects of the nuclear structure can be studied, using  as a series expansion of the electronic wave functions within the nucleus \cite{BW}. Independently, the same series was also used in studies of the penetration effects in the cases of anomalous conversion \cite{sliv,church}. In this way, the  series expansion over the multipole moments $R_2$, $R_4$ $\ldots$ of the nuclear magnetization distribution was developed for HFS in Ref. \cite{NP}.
The leading term is proportional to the square of the rms radius ${R_2}^2$, or the second moment of the distribution of magnetism. Therefore, if two nuclear models have the same $R_2$, they will result in the same HFS in this approximation.  This justifies the common procedure of studies of variations of
$R_2$ by means of the hyperfine anomalies  along  isotopic chains (e. g. \cite{barza,pirs,berlov}).

Herein, we systematize the result in such a way, which keeps
minimum the difference between the results, obtained in the different models. For this purpose, we
will fit the SC radius to the value which results in the same
values  of $w_{1s}$ and $w_{2s}$, as far as possible, to those obtained in the VC model, respectively. For this purpose, we note that the both models only differ by
the penetration matrix element $t^{i}$. Given an $R_c^{VC}$ radius, the  fit of the
$R_c^{SC}$ radius was fulfilled by minimization of the following
form:
\be \chi = \left
|\frac{t_{1s}^{SC}-t_{1s}^{VC}}{t_{1s}^{VC}}\right| +
\left|\frac{t_{2s}^{SC}-t_{2s}^{VC}}{t_{2s}^{VC}}\right| \,. \label{minim}
\ee

       Even in such different models as the ones we use, the coincidence of the HFS values  is achieved up to six decimals.
For the purpose of better comparison to experiment, we added the latest values of QED corrections to the calculated HFS values \cite{shab2}:
$\Delta E_{QED}^{1s}$ = -0.0268 eV, $\Delta E_{QED}^{2s}$ = -0.005 eV, and used the contribution from the electron-electron interactions for the Li-like configuration from Ref. \cite{shab2}, where they were calculated up to the third order of $1/Z$: $\Delta E_{e-e}^{2s}$ = -0.030 eV. The results are presented in Table \ref{tfin}
\begin{table}
\caption{\footnotesize HFS values for the $1s$ and $2s$ states (in eV) calculated with various representative  model radii (in fm). The results are presented for the two values of $\mu$ = 4.1106 and 4.092 nuclear magnetons (see text)}
\begin{center}
\begin{tabular}{||c|c|c||c|c|c|c||}
\hline  \hline
 &&&\multicolumn{2}{c|}{$\mu$=4.1106}&\multicolumn{2}{c||}{$\mu$=4.092} \\
\cline{4-7}
 \raisebox{0.6em}{$R_c^{VC}$} & \raisebox{0.6em}{$R_2$} & \raisebox{0.6em}{$R_c^{SC}$}
  & $W_{1s}$ & $W_{2s}$ & $W_{1s}$ & $W_{2s}$ \\
\hline
  9.1214 & 7.4476 & 7.3703  & 5.06970 & 0.794952 & 5.04663 & 0.791378  \\
8.6214   & 7.0393 & 6.9728 &  5.08041 & 0.796738    &    5.05729 & 0.793156 \\
  8.1214  & 6.6311 & 6.5745  &  5.09087 & 0.798484 &  5.06770 & 0.794894    \\
  7.6214  & 6.2228 &  6.1753 &  5.10106 & 0.800184 & 5.07784 & 0.796586 \\
  7.1214 & 5.8146 & 5.7754 & 5.11092 & 0.801830 & 5.08766  & 0.798224    \\
  6.6214 & 5.4063  & 5.3744 &  5.12042 & 0.803414 & 5.09711 & 0.799801  \\
  6.1214  & 4.9981  & 4.9726 & 5.12949 & 0.804927 & 5.10614 & 0.801308  \\
  5.6214  & 4.5898  & 4.5698 & 5.13809  & 0.806363 & 5.11470 & 0.802737 \\
  \hline    \hline
\end{tabular}
\end{center}
\label{tfin}
\end{table}
for    various representative values of the radii of the models. In the first and third columns, $R_c^{VC}$ and  related $R_c^{SC}$ radii are listed, respectively. The resulting rms radii turn out to be different in the both models. In order to show this, we list the rms $R_2^{VC}$  in the second column. In the SC model, $R_2^{SC} \equiv R_c^{SC}$. For the sake of clarity, the values obtained with $\mu$ = 4.092, are also presented in columns 6 and 7.  In the both cases, the results are in quite satisfactory agreement with the last experimental values of 5.08503(11) and 0.797645(18) eV, although with different rms radii. This difference compensates the variation of the magnetic moment of the nucleus.

\begin{table}
\caption{\footnotesize Calculated $\zeta$ and SD  values for the VC and SC models for the same representative values of the model radii as in Table \ref{tfin}. The SD values are presented in meV }
\begin{center}
\begin{tabular}{||c||c|c|c|c||}
\hline  \hline
$R_c^{VC}$ & $\zeta^{VC}$ & $\zeta^{SC}$ & $\Delta'E$, VC & $\Delta'E$, SC  \\
\hline
  9.1214  & 0.16688  & 0.16688  &  -61.11  &  -61.12  \\
  8.6214  & 0.16688   &0.16689  &  -61.12   & -61.14  \\
  8.1214  & 0.16689   &0.16689  &  -61.14   & -61.15  \\
  7.6214  & 0.16689   &0.16689  &  -61.16    &-61.17  \\
  7.1214  & 0.16689   &0.16690  &  -61.18    &-61.19  \\
  6.6214  & 0.16690   &0.16690  &  -61.20  &  -61.21  \\
  6.1214  & 0.16690   &0.16690  &  -61.22  &  -61.24  \\
  5.6214  & 0.16691   &0.16691  &  -61.25  &  -61.26  \\
\hline  \hline
\end{tabular}
\end{center}
\label{tzeta}
\end{table}

       Proceeding with the $\zeta$ and $\Delta'E$ values, one can see from Eq. (\ref{minim}) that the condition of model independence of the $\zeta$ value
is equivalent to the condition that both of the $t^\nu$ values,
and therefore, both of the $W_{1s}$ and $W_{2s}$ HFS's, might be fitted simultaneously
by different models. This condition is looser than mere
proportionality of the $1s$- and $2s$ wave functions \cite{shab1}, not speaking on that the proportionality is in fact broken by the $e-e$ interactions, QED effects {\it etc}.  Naturally, if the equivalence of the models were
full, the $\zeta$ value would coincide in the both models.
Differences in the $W_i$ values also give rise to the differences
in the $\zeta$-- and $\Delta'E$ values. All these consequences are illustrated in Table \ref{tzeta} for the same representative model radii as in Table \ref{tfin}.
In accordance with what is said previously, the $\zeta$- and $\Delta'E$  values differ from one another much more, than the HFS values.
$\zeta$'s coincide up to fifth decimal, and $\Delta'E$ [��2]values hold up only to the third one. Both values are very sensitive to the only model parameters $R_c^{i}$.

    In finer detail, results of the fit of the $W_i$ values within the framework of the VC model are presented in Table \ref{tcmp}, together with the experimental data. As one can see, the results are not critical to the $\mu$ value: decrease of the latter correlates with adequate decrease of the fitted $R_c$ value without worsening the quality of the fit. A similar fit can be performed, using the SC model. The results obtained in paper \cite{NP} within the framework of the two-parameter magnetic moment method are also presented. They are in good agreement with the present calculations. For comparison, the results  of Refs. \cite{shab2,new} are also listed.
One can see that the latters are in worse agreement with experiment. In contrast, the authors of \cite{new} are quite satisfied by their fit.
Here is the key point to understanding the $^{209}$Bi hyperfine puzzle. As a matter of fact, the authors of \cite{new} compare to experiment not the $W_i$ values themselves, but the SD values instead, which are specially constructed by themselves for this purpose. That such a way is misleading, is explained previously. This delusion leads to underestimated values   of the radii of the nuclear magnetization, as compared to  ours.
\begin{table}
\caption{\footnotesize  Comparison of theoretical results to experiment }
\begin{center}
\begin{tabular}{||c||c|c|c|c|c|c||}
\hline \hline
\raisebox{-0.4em}{Electronic}&\raisebox{-0.4em}{Experiment} &\multicolumn{3}{c|}{$\mu$ = 4.1106}&\multicolumn{2}{c||}
{$\mu$ = 4.092}  \\
\cline{3-7}
\raisebox{0.4em}{state} & \raisebox{0.4em}{\cite{new}} & \cite{NP} & present & \cite{shab2} & present & \cite{new} \\
\hline
$1s$ & 5.08503(11) & 5.0863 & 5.08584 & 5.16138 & 5.08420 & 5.089  \\
$2s$ & 0.797645(18) & 0.7975 & 0.797645 & 0.810230 & 0.797646 & 0.7983  \\
\hline \hline
\end{tabular}
\end{center}
\label{tcmp}
\end{table}

\section{Conclusion}

We performed study of some effects arising in  description of the
Bohr---Weisskopf effect. As expected, the above results  disavow the concept of
the specific differences as a  significant model independent value. These exhibit even stronger sensitivity to the models used than the HFS themselves.
On the other hand, they show that there is no fundamental problem
in  the interpretation of the Bohr---Weisskopf  effect: by mere
fitting the parameters, the effect  can be equally well reproduced by
either of the models within six decimals,
which is quite enough for the present purposes.  At the same time, we have to conclude that such an equivalence
means absence of physical sense in agreement of either of the models with experiment. In the other
words, one cannot conclude that the real distribution is surface-
or volume-like one, based on agreement with experiment, as the
models are mutually exclusive.

    Regarding the calculated values of HFS, they remain in general agreement with  experiment, as in \cite{NP}. This is true with both values of $\mu_n$ = 4.1106 and 4.092, with the corresponding values of the model parameters $R_c^{VC}$ and $R_c^{SC}$.

    Without going into details, let us draw an analogy of the results
obtained above with the proton radius puzzle. The proton rms radius extracted
from the levels in  muonic hydrogen turns out to be different from
that retrieved by means of electron scattering experiment. But we
already saw above that description of the levels does not provide
the rms radius unambiguously. Instead, the value retrieved depends
on the model used.  Cf.  also Refs. \cite{rp1,rp2}.

    An alternative way, based on the two-parameter model,
was proposed in Ref. \cite{NP}. This way allows one to unambiguously retrieve
objective characteristics of the distribution of magnetism inside
the nucleus, such as the second and fourth moments. Model
independence of the values thus obtained has been demonstrated in \cite{NP}. Within this method, the difference obtained above  in the rms values, calculated within the SC and VC models, can be attributed as a manifestation of the truncated terms, containing  $R_4$ and higher moments. Analysis of the results presented in Table \ref{tcmp} suggests that it might be impossible to describe the HFS values for the both levels simultaneously within the framework of an one-parameter model. The two-parameter method of magnetic moments \cite{NP} can. Further research, both experimental and theoretical, is needed in order to better understand the above peculiarities. Specifically, measuring the $2p_{1/2}$ HFS value may be critical to this end \cite{NP}.

\bigskip
\qquad
        The authors would like to express their gratitude to L. F. Vitushkin, D. P. Grechukhin, V. M. Shabaev, I. I. Tupitsin  for  fruitful discussions of the topic and helpful comments.


\end{document}